\documentclass[fleqn,manuscript,prabib]{revtex4}
\usepackage{graphicx}
\def\be{\begin{equation}}
\def\ee#1{\label{#1}\end{equation}}

\begin{document}

\title{Thermodynamics and Kinetic Theory of Relativistic Gases \\ 
in  2-D Cosmological Models }
\author{G. M. Kremer\thanks{kremer@fisica.ufpr.br}  $\,$ and 
F. P. Devecchi\thanks{devecchi@fisica.ufpr.br}\\
Departamento de F\'\i sica, Universidade Federal do Paran\'a\\
Caixa Postal 19044, 81531-990, Curitiba, Brazil
}
\date{}
\begin{abstract}
A kinetic theory of relativistic gases in a two-dimensional space is developed 
in order to obtain the equilibrium distribution function and the expressions 
for the 
fields of energy per particle, pressure, entropy per particle and heat 
capacities in equilibrium.  Furthermore, by using the method of 
Chapman and Enskog for a kinetic model of the Boltzmann equation 
the non-equilibrium energy-momentum tensor 
and the  entropy production rate are determined for a universe described by a 
two-dimensional
Robertson-Walker metric. 
The solutions of the gravitational field equations that consider the 
non-equilibrium energy-momentum tensor - associated with the coefficient of 
bulk viscosity - show that opposed to the 
four-dimensional case, the cosmic scale factor attains a maximum value at a 
finite time decreasing to a "big crunch" and that there exists a solution of 
the gravitational field equations corresponding to a "false vacuum". The
evolution of the fields of pressure, energy density and entropy production 
rate with the time is also discussed.

\end{abstract}
\maketitle
\noindent
PACS: 51.10.+y; 98.80.-k; 47.75.+f
\section {Introduction} 

The combination of general relativity with the kinetic theory of gases
is remarkably useful to construct cosmological models~\cite{Bern}. In
these formulations
the  cosmic sources of gravitational interactions are represented by the
energy-momentum tensor of a  fluid; in addition
we have the hypothesis of homogeneity and isotropy in the form of the
well-known Robertson-Walker metric~\cite{Wein1}. Although these theories
 have explained  several important features of our universe  fundamental
 questions still remain to be 
answered~\cite{Olive}. 

Models in lower dimensions offer interesting results that, if properly
analyzed, can be used to gain insight in the realistic formulations.
Two-dimensional (2-D) gravity models have been under intensive
investigation
during the last
two decades~\cite{Teit,Jackiw,Pol,Stro,Mann,Mann1}. 
The old problem of quantum gravity, black holes
physics and string dynamics were tested in
these theories. In particular Teitelboim~\cite{Teit} and
Jackiw~\cite{Jackiw} proposed
a consistent model in two
dimensions analogous of general relativity. As imediate results,
among
 others~\cite{Jackiw,Mann},  this model
offer a consistent Newtonian limit, gravitational collapse solutions
that are basically a 2-D Schwarzschild analogue and cosmological
models based in a 2-D Robertson-Walker metric. 

For cosmological applications, a  refinement in the construction of
 these models can be  obtained by considering a non-equilibrium
scenario, including a bulk viscosity term in the energy-momentum tensor 
(for a review on viscous cosmology up to 1990 one is referred to 
Gron~\cite{Gron}).  
In the four-dimensional case the inclusion of this term to analyze the 
evolution of 
the cosmic scale factor with the time was done by Murphy~\cite{Mur} who has 
found   
a solution that corresponds only to an expansion. 
Other models were based on the coupling of the Einstein field equations with 
the balance equations of extended thermodynamics~\cite{MR} (also known as 
causal or second-order thermodynamic theory) and among others we cite the 
works of
Belinski\v i et al~\cite{Bel}, Zimdahl~\cite{Zim} and Di Prisco et 
al~\cite{Di}.

In this work we develop a kinetic theory of relativistic gases in a 
two-dimensional space.
The balance laws for the particle flow, 
energy-momentum tensor and entropy flow are obtained from the
Boltzmann equation. We find also the equilibrium distribution 
function and the expressions for the fields of energy per particle, pressure, 
entropy per particle, 
enthalpy per particle and heat capacities in equilibrium in a two-dimensional 
space. 
Moreover, by using the method of 
Chapman and Enskog for the kinetic model of the Boltzmann equation 
proposed by Anderson and Witting~\cite{AW} we calculate 
the bulk viscosity and the entropy production rate.
We apply the ideas of Murphy~\cite{Mur} to the 2-D  gravitational field 
equations
and  we show that opposed to the four-dimensional case the cosmic scale 
factor attains a maximum value at a finite time decreasing to a "big crunch" 
and that there exists a solution of the gravitational field equations 
corresponding to a "false vacuum". The difference
between the solutions in the four- and two-dimensional cases is due to the
fact that  the relationship between the metric tensor and the sources in 
the 2-D case is modified because the Einstein field equations give no 
dynamics for the 2-D case.

The manuscript is structured as follows. In Section II we introduce 
the two-dimensional Robertson-Walker metric. The kinetic theory of 
relativistic gases in 2-D 
is developed in Section III. In Section IV we introduce the gravitational 
equations of motion in the 2-D case and in Section V
we search for the solutions of the gravitational field equations. Finally,
in Section VI we discuss the solutions that came out from the  gravitational 
field equations.

\section{Robertson-Walker Metric}

One fundamental feature of 2-D cosmological models is that they show
considerably less mathematical complexity and at the same time they
preserve
the physical principles that are used to construct their four-dimensional
counterparts. One impressive result was that in 2-D models the 
quantization of the 
gravitational field is  consistent~\cite{Jackiw,Pol}, opening
the possibility of quantum
cosmological models for the very early universe. These results
in the 2-D theories are of relevance to
include new ingredients in the ``realistic'' versions~\cite{Mann,Mann1}.

As it is well known, the so-called cosmological
principle is based on the assumption that the
universe is spatially homogeneous and isotropic. The metric
that describes
such kind of universe, known as the
Robertson-Walker metric, has  the
following form in a two-dimensional Riemannian space characterized
by the metric tensor $g^{\mu\nu}$ with signature $(+ \,-)$~\cite{Mann,Mann1}
\be 
ds^2=(cdt)^2-\kappa(t)^2{(dr)^2\over 1-\varepsilon r^2}.
\ee{RW1}
In the above equation  
$\kappa(t)$ - the so-called cosmic scale factor - is an unknown function of 
the time and has dimension of length, while  
$r$ is a dimensionless quantity. If we introduce a new variable 
$x=\arcsin(\sqrt{\varepsilon}r)/\sqrt{\varepsilon}$ the equation (\ref{RW1})
reduces to
\be 
ds^2=(cdt)^2-\kappa(t)^2(dx)^2.
\ee{RW1a}

The components  and the determinant $g=-\det((g_{\mu\nu}))$
of the metric tensor $g_{\mu\nu}$ for the
Robertson-Walker metric (\ref{RW1})
with respect to the coordinates $(x^\mu)=(ct, x)$ are 
\be
g_{00}=g^{00}=1,\quad g_{11}=-\kappa^2
={1\over g^{11}},\quad g=\kappa^2
\ee{RW15}
The corresponding non-zero Christoffel symbols
 read 
\be
\Gamma_{11}^0=\dot\kappa\kappa,\quad
\Gamma_{01}^1={\dot\kappa\over \kappa}.
\ee{RW17}
where the dot denotes the derivative with respect to the coordinate $x^0=ct$.
Once the Christoffel symbols are known the
non-vanishing
components of the Ricci tensor 
$R_{\mu\nu}=R^\tau_{\mu\tau\nu}$ and the curvature scalar
$R=g^{\mu\nu}R_{\mu\nu}$ can be calculated and it follows
\be
R_{00}={\ddot \kappa\over\kappa},\qquad
R_{11}=-\ddot\kappa\kappa,
\qquad
R=2{\ddot\kappa\over\kappa}.
\ee{RW21}

\section{Kinetic Theory of Relativistic Gases in 2-D}

\subsection{Boltzmann equation and fields in equilibrium}

We consider a relativistic ideal gas with particles of rest mass
$m$  which is described in the phase space by
the one-particle distribution function $f(x^\mu, p^\mu)$. 
The momentum  $(p^\mu)=(p^0, p)$ has a
constraint of constant length 
$g_{\mu\nu}p^\mu p^\mu=m^2c^2$ so that 
$f(x^\mu, p^\mu)\equiv f(x, p, t)$. 

The evolution equation of the one-particle distribution function in the
phase space is described by the Boltzmann equation, which in the 
presence of a gravitational field  is given by~\cite{CK1,Cher}
\be
 p^\mu{\partial f\over
\partial x^\mu}-\Gamma_{\mu\nu}^1p^\mu p^\nu{\partial f\over\partial p}
=-{U^\mu p_\mu\over c^2\tau}(f-f^{(0)}),
\ee{1}
where $\Gamma^1_{\mu\nu}$ is the affine connection.
In the right-hand side of the above equation we have replaced the collision
term of the Boltzmann equation by the model equation proposed by Anderson and 
Witting~\cite{AW}, which refers to a relativistic gas in the Landau and 
Lifshitz description~\cite{LL}. Further
 $\tau$ is a characteristic time of order
of the time between collisions, $f^{(0)}$ is the equilibrium distribution 
function and $U^\mu$  is the two-velocity such that
$U^\mu U_{\mu}=c^2$.

The first moments of the one-particle distribution function are: 
the particle flow
$N^{\mu}$ and the energy-momentum tensor $T^{\mu\nu}$ which are defined by
\be
N^{\mu}=c\int  p^{\mu}f\,\sqrt{g} {dp\over p_0},\qquad
T^{\mu\nu}=c\int
p^{\mu} p^{\nu} f\,\sqrt{g}
{dp\over p_0}.
\ee{6}
Furthermore, the entropy flow $S^\mu$ is defined by
\be
S^{\mu}=-kc\int p^{\mu}f\,\ln f\,\sqrt{g}{dp\over p_0},
\ee{9}
where $k$ is the Boltzmann constant.

The balance equations for the particle flow  $N^{\mu}$, 
energy-momentum tensor  $T^{\mu\nu}$ and entropy flux $S^\mu$ can be
obtained from the Boltzmann equation (\ref{1}) and read
\be
{N^{\mu}}_{;\mu}=0,\qquad
{T^{\mu\nu}}_{;\mu}=0,\qquad
{S^{\mu}}_{;\mu}=\varsigma\geq 0.
\ee{7}
Above $\varsigma$ denotes the entropy  
production rate defined through
\be
\varsigma={kU^\mu\over c\tau}\int p_\mu f^{(0)} \left({f\over 
f^{(0)}}-1\right)\ln {f\over 
f^{(0)}}\sqrt{g} {dp\over p_0}.
\ee{10}

In the Landau and Lifshitz description~\cite{LL} the particle flow and the 
energy-momentum tensor are decomposed according to
\be
N^\mu=nU^\mu+{\cal J}^\mu,
\ee{W2a}
\be
T^{\mu\nu}=p^{\langle\mu\nu\rangle}-(P+\varpi)\Delta^{\mu\nu}
+{en\over c^2}U^\mu U^\nu,
\ee{W2}
where $n$ is the particle number density, $\cal J^\mu$ the non-equilibrium
part of the particle flow, $p^{\langle\mu\nu\rangle}$ the pressure
deviator, i.e., the traceless part of the pressure tensor, 
$P$ the hydrostatic pressure, $\varpi$ the dynamic pressure, i.e., the 
non-equilibrium part of the trace of the energy-momentum tensor, $e$ the 
internal energy per particle and $\Delta^{\mu\nu}$ the projector  defined 
by
\be 
\Delta^{\mu\nu}=g^{\mu\nu}-{1\over c^2}U^\mu U^\nu,
\quad \hbox{such that}\quad \Delta^{\mu\nu}U_{\nu}=0,\quad
\Delta^{\mu\nu}\Delta_\nu^\sigma=\Delta^{\mu\sigma}, \quad
\Delta_{\mu\nu}\Delta^{\mu\nu}=1.
\ee{W3}
The the non-equilibrium
part of the particle flow $\cal J^\mu$ and the pressure
deviator $p^{\langle\mu\nu\rangle}$ are perpendicular to $U^\mu$.

One can also decompose the entropy flow as
\be 
S^\mu=nsU^\mu+\phi^\mu
\ee{11a}
where $s$ denotes  the entropy per particle and $\phi^\mu$ the entropy flux 
which is perpendicular to $U^\mu$.

The equilibrium distribution function in a two-dimensional space, which 
is the so-called Maxwell-J\"uttner distribution function, can be written as
\be
f^{(0)}={n\over
2mc K_1(\zeta)}
e^{-{1\over kT}U^{\mu}p_{\mu}},\qquad\hbox{where}\qquad
\zeta={mc^2\over kT},
\ee{35}
where $T$ denotes the absolute temperature.
The parameter $\zeta$ represents the ratio
between the rest energy of a particle and the thermal energy of the gas.
Moreover, $K_n(\zeta)$ denotes the modified Bessel function of 
second kind (see for example~\cite{AS})
\be
K_n(\zeta)=\left({\zeta\over 2}\right)^n
{\Gamma({1\over 2})\over \Gamma(n+{1\over 2})}
\int_1^\infty e^{-\zeta y}(y^2-1)^{n-{1\over 2}}dy.
\ee{22}

Once the equilibrium distribution function is known, one can obtain
the following expressions for the fields of energy per particle,
pressure  and entropy per particle in equilibrium:
\be
e=mc^2\left[
{K_2(\zeta)\over K_1(\zeta)} -{1\over \zeta}\right],\qquad P=nkT,\qquad
s_{ E}=k\left\{{e\over kT}-\ln\left[
{n\over 2mc K_1(\zeta)}\right]\right\}.
\ee{36}

The thermodynamic quantities enthalpy per particle $h$  and heat capacities
per particle at constant volume $c_v$ and at constant pressure $c_p$ 
follows from its definitions by using (\ref{36}), yielding
\be
h=e+{P\over n}=mc^2{K_2(\zeta)\over K_1(\zeta)},\quad
c_{ v}=\left(
{\partial e\over \partial T}\right)_{ v}=k\left[\zeta^2
+3\zeta{K_2(\zeta)\over K_1(\zeta)}-\zeta^2\left({K_2(\zeta)\over K_1(\zeta)}
\right)^2-1\right],
\ee{37}
\be
c_{ p}=\left(
{\partial h\over \partial T}\right)_{ p}=k\left[\zeta^2
+3\zeta{K_2(\zeta)\over K_1(\zeta)}-\zeta^2\left({K_2(\zeta)\over K_1(\zeta)}
\right)^2\right]=c_v+k.
\ee{38}

The above thermodynamic fields in the non-relativistic limiting case 
where $\zeta\gg1$, i.e., for low temperatures,  read
\be
e\approx mc^2+{kT\over 2}\left[1+{3\over 4\zeta}+\dots\right],
\qquad
s_E\approx k\left[{1\over 2}+\ln\sqrt{2\pi mk}+\ln{\sqrt{T}\over n}
+{1\over \zeta}+\dots\right],
\ee{39}
\be
h\approx mc^2+{3kT\over 2}\left[1+{3\over 4\zeta}+\dots\right],\quad
c_v\approx {k\over 2}\left[1+{3\over 2\zeta}-{9\over 4\zeta^2}+\dots\right].
\ee{40}

In the ultra-relativistic limiting case where $\zeta\ll1$, i.e., for high 
temperatures and/or for very small rest mass,  we have that 
the leading term of each thermodynamic field is given by 
\be
e\approx kT={P\over n}, \qquad
s_E\approx k\left[1+\ln{2kT\over nc}\right], \qquad h\approx  2kT,\qquad
c_v\approx 3k.
\ee{41}

\subsection{Dynamic Pressure in a  Homogeneous and Isotropic Universe}

We shall determine in this section the dynamic
pressure and the entropy 
production rate in a spatially homogeneous and 
isotropic universe. In the four-dimensional case these
topics were  discussed by Weinberg~\cite{Wein}
within the framework of a phenomenological 
theory  and by  Bernstein~\cite{Bern} 
within the framework of a kinetic 
theory of gases.
Without loss of generality, we shall use here the  Anderson and Witting model 
of the Boltzmann equation (\ref{1}) in order to simplify the calculations.

We begin by neglecting the space gradients - since we are dealing with a 
spatially  homogeneous and isotropic universe - and by considering a 
comoving frame where $(U^\mu)=(c,  0)$ in a two-dimensional space. Hence 
it follows that the Boltzmann equation (\ref{1}) reduces to
\be
{\partial f\over\partial x^0}-2{\dot\kappa\over\kappa}p{\partial f\over
\partial p}=-{1\over c\tau}(f-f^{ (0)}).
\ee{12.65}

We use now the Chapman and Enskog method (see for example~\cite{CK1}) and 
search for a solution of the
Boltzmann equation (\ref{12.65}) of the form 
\be 
f=f^{(0)}(1+\phi),
\ee{12.66} 
where $f^{(0)}$ is the Maxwell-J\"uttner distribution function 
(\ref{35}) and $f^{(0)}\phi$ is the  deviation from equilibrium of 
the one-particle distribution function. 
We insert the above representation into the Boltzmann equation 
(\ref{12.65}) and by taking into account only the derivatives
of the Maxwell-J\"uttner distribution function 
we get that the deviation $\phi$ is given by
$$
\phi=-c\tau\left({\partial \ln f^{(0)}\over\partial x^0}-
2{\dot\kappa\over\kappa}p{\partial \ln f^{(0)}\over \partial p}\right)
$$
\be
=-c\tau 
\left[{\dot n\over n}+\left(1-\zeta{K_2\over K_1}\right){\dot T\over T}
+{c\over kT}p_0{\dot T\over T}+{c\over kT}{\dot\kappa\over\kappa}{p^2
\over p_0}\right].
\ee{12.70}
For the elimination of $\dot n$ and $\dot T$ from (\ref{12.70}) 
 we use the balance 
equations of the particle flow and 
energy-momentum tensor of a non-viscous and non-heat-conducting relativistic
gas whose constitutive equations read:
\be 
N^\mu=nU^\mu,\qquad
T^{\mu\nu}=(ne+P){U^\mu U^\nu\over c^2}-Pg^{\mu\nu}.
\ee{12.73}
Insertion of the above representations into the balance equations 
${N^\mu}_{;\mu}=0$ and ${T^{\mu\nu}}_{;\nu}=0$ lead to
\be
{1\over \sqrt{g}}{\partial \over\partial x^\mu}\left(\sqrt{g}nU^\mu\right)=0,
\ee{12.74}
\be
-{\partial P\over\partial x^\nu}g^{\mu\nu}+{1\over \sqrt{g}}
{\partial\over\partial x^\nu}\left[\sqrt{g}(ne+P) {U^\mu U^\nu\over c^2}
\right]+(ne+P)\Gamma_{\lambda\nu}^\mu {U^\lambda U^\nu\over c^2}=0.
\ee{12.75}
In a comoving frame (\ref{12.74}) becomes
\be
{\dot n\over n}=-{\dot\kappa\over \kappa},
\ee{12.76}
while the spatial components of (\ref{12.75}) are identically satisfied due to
the constraint that all quantities $n, P, e$ and $\kappa$ are only functions 
of the time coordinate. The temporal component of (\ref{12.75}) can be 
written as 
\be
{\dot T\over T}=-{k\over c_v}{\dot\kappa\over \kappa},
\ee{12.77}
where the heat capacity per particle at constant volume $c_v$ is given by 
(\ref{37})$_2$.
Equations (\ref{12.76}) and (\ref{12.77}) are used to eliminate $\dot n$ and 
$\dot T$ from (\ref{12.70}), yielding
\be
\phi=c\tau
\left[1+\left(1-\zeta{K_2\over K_1}\right){k\over c_v}
+{c\over c_vT}p_0-{c\over kT}{p^2
\over p_0}\right]{\dot\kappa\over\kappa}.
\ee{12.70a}

Once  the non-equilibrium distribution
function (\ref{12.66}) is known it is possible to 
calculate the projection of the energy-momentum
tensor in a comoving frame 
which corresponds to the sum of the hydrostatic pressure 
with the 
dynamic pressure, i.e.,
\be
P+\varpi=-\Delta_{\mu\nu}T^{\mu\nu}=-\left(g_{\mu\nu}-
{U_\mu U_\nu\over c^2}\right)c\int p^\mu p^\nu f\sqrt{g}{dp\over p_0}
=c\kappa^3\int p^2f
{dp\over p_0}.
\ee{12.71}
We insert (\ref{12.66}) together with
(\ref{12.70a}) into (\ref{12.71}) and integrate the resulting equation,
yielding
\be
\varpi=-\eta
c{\dot\kappa\over\kappa},\qquad\hbox{where}\qquad \eta=
-P\tau\left\{\zeta^2\left[
{K_2\over \zeta K_1}-1+{{\rm Ki_1}\over K_1}
\right]-1-{1\over c_v}\right\}.
\ee{12.79}
In the above equation
${\rm Ki}_1$ denotes the integral for the modified
Bessel functions\index{modified Bessel function}
(see Abramowitz and Stegun~\cite{AS} p. 483):
\be
{\rm Ki}_n(\zeta)=\int_\zeta^\infty {\rm Ki}_{n-1}(t)dt=
\int_0^\infty {e^{-\zeta\cosh t}\over \cosh^nt}dt.
\ee{CT11}

Hence we have identified the coefficient of proportionality between 
$\varpi$ and $c{\dot\kappa/\kappa}$ as the bulk
viscosity $\eta$. If we compare (\ref{12.79}) with the constitutive equation
for the dynamic pressure, given in terms of the 
divergence of the two-velocity,
i.e., $\varpi=-\eta{U^\mu}_{;\mu}$, we infer that here $c\dot\kappa/\kappa$ 
plays the same role as ${U^\mu}_{;\mu}$. Furthermore due to the fact that
 the bulk viscosity is a positive quantity the 
dynamic pressure decreases 
when the universe is expanding ($\dot\kappa>0$) while it 
increases when the 
universe is contracting ($\dot\kappa<0$).

As in the four-dimensional case the coefficient of bulk viscosity vanishes in 
the 
non-relativistic and ultra-relativistic limiting cases. This can be seen 
from Figure 1 where  the coefficient of bulk viscosity $\eta\sigma/
\sqrt{kTm}$ is plotted versus the parameter $\zeta=mc^2/(kT)$. Here we have
chosen the following expression for the characteristic time~\cite{CK3}  
\be
\tau={1\over n\sigma v_s},\qquad\hbox{with}\qquad v_s=\sqrt{{c^2c_pkT\over
c_v h}},
\ee{A}
where $\sigma$ is a differential cross-section and $v_s$ the adiabatic sound
speed.

If we consider that the distribution function 
is given by (\ref{12.70}), i.e.,
$f=f^{(0)}(1+\phi)$, we can use  the approximation $\ln (1+\phi)\approx 
\phi$ valid for $\vert\phi\vert\ll 1$ in order 
to write the entropy production rate  (\ref{10}) 
in a comoving frame as
\be
\varsigma={k\over \tau}\int f^{(0)}\phi^2\sqrt{g}dp.
\ee{12.81}
We insert now (\ref{12.66})  together with (\ref{12.70a}) into  (\ref{12.81})
and get by integrating the resulting equation
\be
\varsigma={\eta c^2\over T}\left({\dot\kappa\over\kappa}\right)^2.
\ee{12.82}
Hence the entropy production rate is connected 
with the bulk viscosity. 
Weinberg~\cite{Wein} has derived a similar formula in a four-dimensional space 
by using a phenomenological
theory and has also shown that the bulk viscosity alone 
could not explain
the high entropy of the present microwave background radiation. For more 
details one is referred to Weinberg~\cite{Wein,Wein1}.

\section{Gravitational Equations of Motion}

Given the sources of our universe it is possible to obtain the equations
of motion for the gravitational field in the whole space-time. The starting
point is of course, the Einstein theory of gravitation.

As far as we are working in 2-D some fundamental problems appear. In fact, the
main point is that the Einstein action in 2-D furnishes no dynamics.
In other words the usual left-hand side of Einstein equations of motion that
follows by using the Hamilton variational principle  
\begin{equation}
R_{\mu \nu}- \frac{1}{2}g_{\mu \nu }R \equiv 0,
\end{equation}
are in fact an {\it identity} in 2-D. This is related to the
gauge invariances of gravitation in 2-D: space-time diffeomorphisms and
(local) conformal transformations.   With this in mind Teitelboim~\cite{Teit} 
and 
Jackiw~\cite{Jackiw} proposed as 2-D action\footnote{From this section  
on units have been chosen so that $G=c=k=1$.}
\begin{equation}
S= \int d^2x \sqrt{-g}\left\{N(x)\left[R(x)+8\pi
T^\mu_\mu(x)\right]\right\},
\end{equation}
where $T^\mu_\mu(x)$ is the trace of the energy-momentum
tensor. Above we have not taken into
account the term that refers to the cosmological constant. 
Using the variational 
principle for the auxiliary field  $N(x)$ the equation of motion that 
follow is
\be
R (x) = -8\pi T^\mu_\mu(x),
\ee{RW21a}
together with the conservation law ${T^{\mu\nu}}_{;\nu} =0$. Equation 
(\ref{RW21a}) relates
the curvature scalar with the trace of the energy-momentum tensor.

\section{Solution of Gravitational Field Equations}

The gravitational field equations are obtained from (\ref{RW21a})
together with  ${T^{\mu\nu}}_{;\nu} =0$  and by taking into account the 
constitutive equation for the energy-momentum
tensor $T^{\mu\nu}=\epsilon U^\mu U^\nu-(P+\varpi)\Delta^{\mu\nu}$ where
$\epsilon=ne$ is the energy density. Hence it follows that
\be
\ddot\kappa=-4\pi(\epsilon-P-\varpi)\kappa,\qquad
\dot \kappa(\epsilon+P+\varpi)+\kappa\dot\epsilon=0.
\ee{e1}
Since $\varpi=-\eta\dot\kappa/\kappa$, the above system is closed if we can 
relate the pressure $P$ and the coefficient of bulk
viscosity $\eta$ to the energy density $\epsilon$. Here we follow 
Murphy~\cite{Mur} and assume a barotropic equation of state for the 
pressure and a linear relationship between the coefficient of bulk 
viscosity and the energy density, so that
\be
P=(\gamma-1)\epsilon,\qquad \eta=\alpha\epsilon.
\ee{e2}
Above $\alpha$ is a constant and $\gamma$ may range from 
$0\leq\gamma\leq2$. The value $\gamma=1$  refers to a pure matter (dust), 
$\gamma=2$ to a pure radiation and $\gamma<1$ to a "false vacuum".

If we introduce the Hubble parameter $H=\dot \kappa/\kappa$ the field 
equations (\ref{e1}) can be written thanks to (\ref{e2}) as
\be
\dot H+H^2=4\pi\epsilon(\gamma-2-\alpha H),\qquad
 H(\gamma-\alpha H)\epsilon+\dot \epsilon=0.
\ee{e3}
The elimination of the energy density of the two above equations leads to 
an equation for the Hubble parameter that reads
\be
(\ddot H+2H\dot H)(\gamma-2-\alpha H)+\alpha \dot H(\dot H+H^2)
+H(\gamma-2-\alpha H)(\gamma-\alpha H)(\dot H+H^2)=0.
\ee{e4}

Let us search for a solution of (\ref{e4}) by considering $H=H_*=$constant. 
In this case we get that (\ref{e4}) reduces to 
$(\gamma-2-\alpha H_*)(\gamma-\alpha H_*)H_*^3=0$ and apart from 
the trivial solution $H_*=0$ we have two other
solutions, namely: i) $\alpha=\gamma/H_*$ and ii) $\alpha=(\gamma-2)/H_*$.  
Only the first solution is physically possible since 
$0\leq\gamma\leq2$.

We proceed to analyze the two energy conditions~\cite{Haw}, namely the weak
energy condition which dictates that the energy density is a semi-positive
quantity, i.e., $\epsilon\geq0$ and the strong energy condition which
imposes that the  inequality $\epsilon+P+\varpi\geq0$ must hold
implying that $H/H_*\leq 1$.

We are now ready to determine the solutions of the gravitational field 
equations.
For that end we write the system of equations (\ref{e1}) and (\ref{e3}) as
\be
(\gamma-1){\ddot\kappa\over \kappa}=4\pi P\left
(\gamma-2-\gamma{\dot\kappa\over\kappa}\right),\qquad
\gamma{\dot \kappa\over\kappa}\left(1-{\dot\kappa\over\kappa}\right)
+{\dot P\over P}=0,
\ee{e5}
\be
(\gamma-1)(\dot H+H^2)=4\pi P(\gamma-2-\gamma H),\qquad
 H\gamma(1- H)P+\dot P=0,
\ee{e6}
respectively, where the cosmic scale factor, the Hubble parameter, 
the pressure and the time are 
now taken as dimensionless quantities with respect to $H_*$.
>From the first system of equations (\ref{e5}) one may determine
the cosmic scale factor $\kappa(t)$ and the pressure $P(t)$ while
from the second one (\ref{e6}) the Hubble parameter $H(t)$ can be obtained.
We have chosen two values for $\gamma$, one of them $\gamma=0.5$ correspond 
to a negative
pressure ("false vacuum") while the other $\gamma=1.5$ implies a
positive pressure.
For the solutions of the two systems of equations (\ref{e5}) and (\ref{e6}) 
the following initial conditions were taken into account:
\be\cases{\hbox{for}\quad \gamma=0.5,\quad \hbox{then}\quad H(0)=0.5,\quad 
\kappa(0)=1,\quad P(0)=-1,\cr
\hbox{for}\quad\gamma=1.5,\quad \hbox{then}\quad H(0)=0.5,\quad \kappa(0)=1,
\quad P(0)=1.}
\ee{e7} 
In Figures 2 through 5 it is shown the evolution with respect to the 
time of the cosmic scale factor, the
Hubble parameter, the pressure and the energy density which follow from the 
systems of equations (\ref{e5}) and (\ref{e6}) by taking into account the 
initial conditions (\ref{e7}) and the barotropic equation of state.

Another important quantity which can be plotted versus the time is the 
entropy production rate.  For the cases where $\gamma>1$ the dimensionless 
expression 
for entropy production rate (\ref{12.82}) can be written as
\be
\varsigma={\gamma H^2\over(\gamma-1)\kappa}.
\ee{e8}
In order to obtain the above equation we have used the equation of state
$P=nT$ and the solution of the continuity equation
(\ref{12.76}) which reads $n\propto1/\kappa$. In Figure 6 we show the 
evolution of the entropy production rate with respect to the time.
\section{Discussion of the Results}

We proceed to discuss the results obtained in the last section. First we call
attention to the fact that the period where this theory can be applied 
is the one where there exists an interaction between radiation and matter.
The reason is that in the  period of pure radiation the dynamic pressure 
vanishes while in the period of pure matter (dust) there is no interaction at 
all. 
Furthermore the solutions are not valid for all values of the time, since in 
some period
the radiation decouples from matter and we have no more the interaction 
between 
the radiation and matter which implies a vanishing dynamic pressure.

In Figure 2 it is plotted the cosmic scale factor as function of the time. 
We have chosen that at the beginning of this period ($t=0$ by adjusting clocks)
 the cosmic scale factor is different from zero. We infer from this figure 
that the cosmic scale factor has a maximum at $t\approx0.0243$ for 
$\gamma=1.5$ and at $t\approx0.0123$ for $\gamma=0.5$. From these points on 
the cosmic scale factor decreases and tends to zero, i.e., it goes to 
a "big crunch". It is
noteworthy to call attention to the fact that the corresponding solution in 
the four-dimensional case~\cite{Mur} neither have this behavior for the 
cosmic scale factor
nor admit a "false vacuum" solution. As was previously pointed out the 
difference
between the solutions in the four- and two-dimensional cases is due to the
fact that  the relationship between the metric tensor and the sources in 
the 2-D case is modified because the Einstein field equations give no 
dynamics for the 2-D case.

The Hubble parameter as a function of the time is shown in Figure 3. 
For both values of 
$\gamma$ the Hubble parameter decrease, attain a zero value for a time 
which correspond
to the maximum of the cosmic scale factor and assume negatives values.

In Figures 4 and 5 the evolution of the pressure and of the energy density 
are represented 
as functions of the time, respectively. Both functions decrease with the 
time and attain their minimum values at the times where the cosmic scale 
factor has its maximum value. Since in this
theory there exists no mechanism that could increase the pressure and the 
energy density, we infer that the solutions for $t{^>_\sim}0.0243$ when 
$\gamma=1.5$ and 
for $t{^>_\sim}0.0123$ when $\gamma=0.5$ are not physically possible. 
The same conclusion can be drawn
out from Figure 6 where the evolution of the entropy production rate is 
plotted as a function of the time. From this figure we note that the 
entropy production rate decreases with the time and attains its minimum 
($\varsigma=0$) when the cosmic scale factor reach its maximum value.
At this point the entropy per particle $s$ attains its maximum value 
since $nDs=\varsigma\geq0$. There exists no mechanism in this theory  
that could increase the entropy density rate from its minimum value 
with a corresponding  decrease of the entropy  per particle from its 
maximum value.

\newpage
\section*{Figure Captions}

\noindent
Figure 1: Volume viscosity $\eta\sigma/\sqrt{kTm}$ vs. $\zeta$.
\vskip1cm

\noindent
Figure 2: Cosmic scale factor $\kappa(t)$ vs. time $t$.
\vskip1cm

\noindent
Figure 3: Hubble parameter $H(t)$ vs. time $t$.
\vskip1cm

\noindent
Figure 4: Pressure $P(t)$ vs. time $t$.
\vskip1cm

\noindent
Figure 5: Energy density $\epsilon(t)$ vs. time $t$.
\vskip1cm

\noindent
Figure 6: Entropy production rate $\varsigma(t)$ vs. time $t$.

\newpage
\begin{figure}
\begin{center}
\includegraphics[width=11cm]{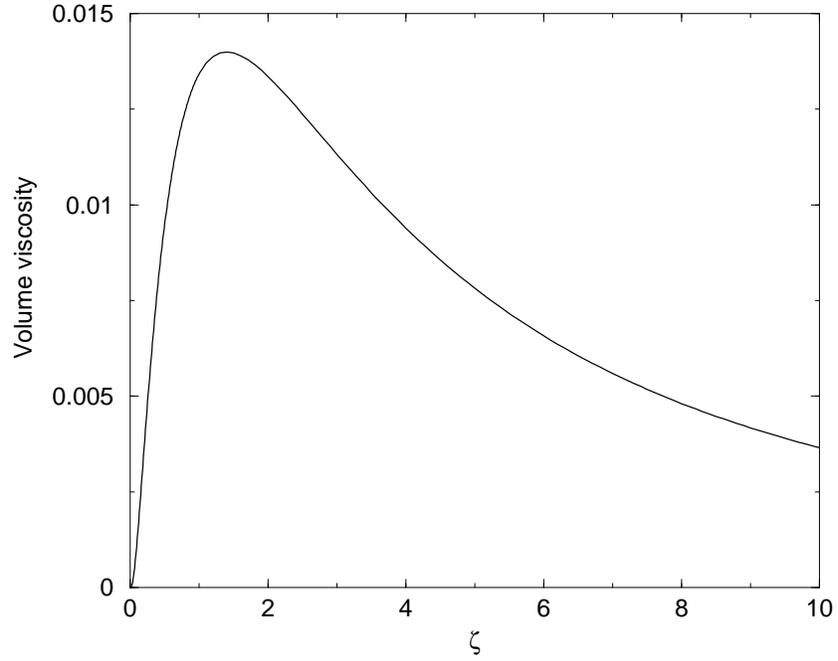}
\caption{Volume viscosity $\eta\sigma/\sqrt{kTm}$ vs. $\zeta$.}
\end{center}
\end{figure}

\begin{figure}
\begin{center}
\includegraphics[width=11cm]{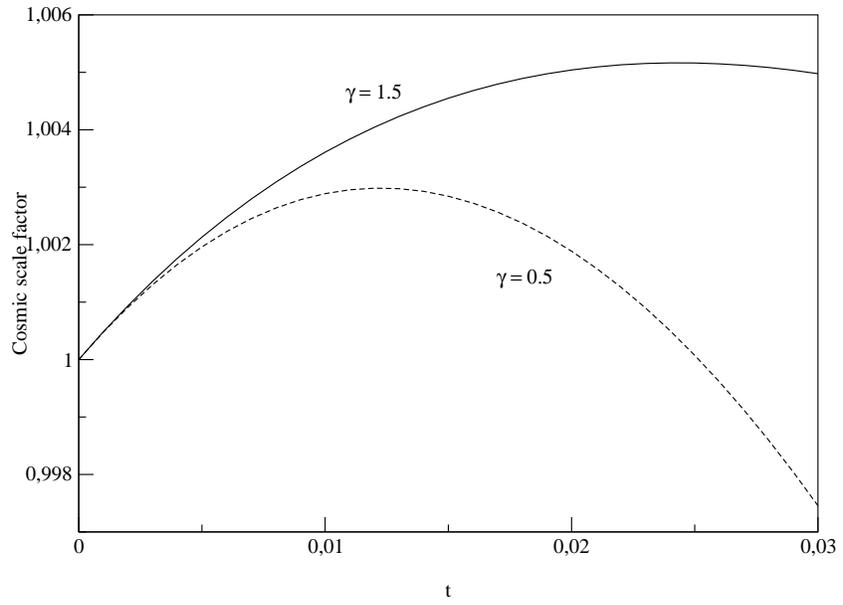}
\caption{Cosmic scale factor $\kappa(t)$ vs. time $t$.}
\end{center}
\end{figure}

\begin{figure}
\begin{center}
\includegraphics[width=11cm]{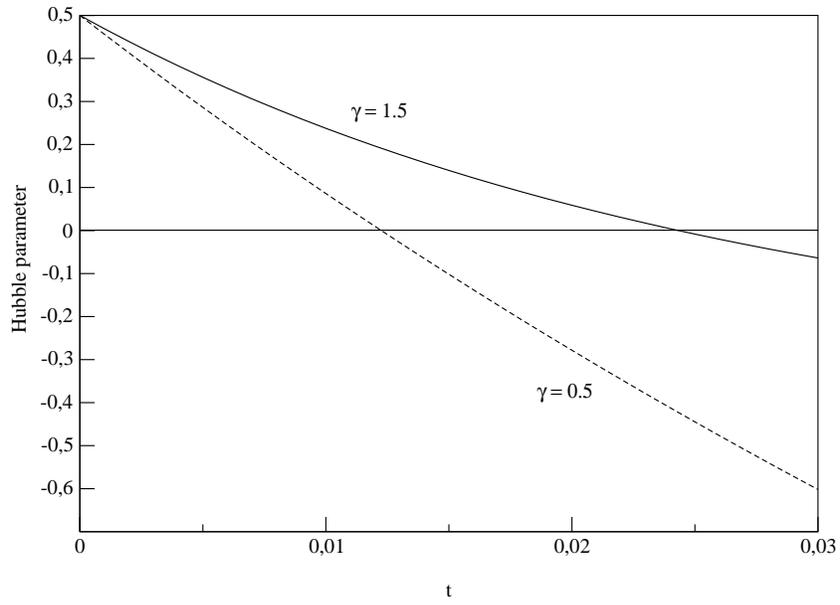}
\caption{
Hubble parameter $H(t)$ vs. time $t$.}
\end{center}
\end{figure}

\begin{figure}
\begin{center}
\includegraphics[width=11cm]{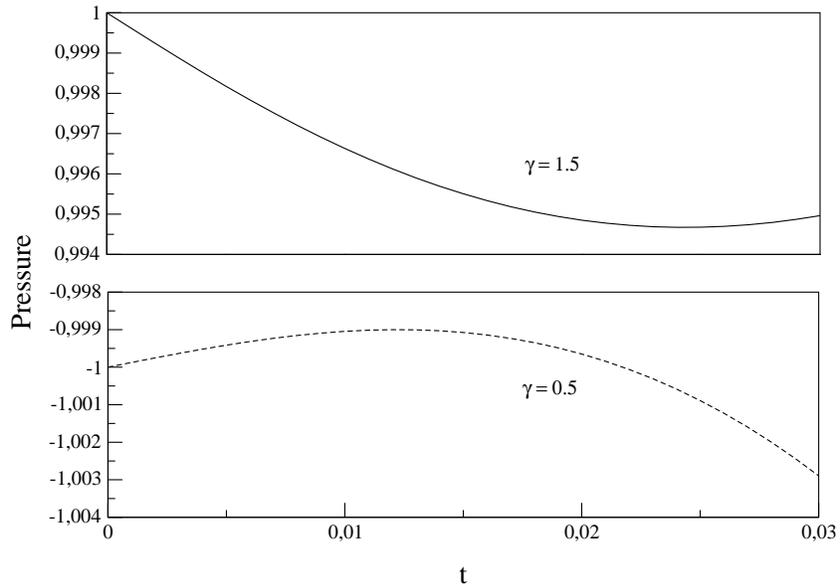}
\caption{Pressure $P(t)$ vs. time $t$.}
\end{center}
\end{figure}

\begin{figure}
\begin{center}
\includegraphics[width=11cm]{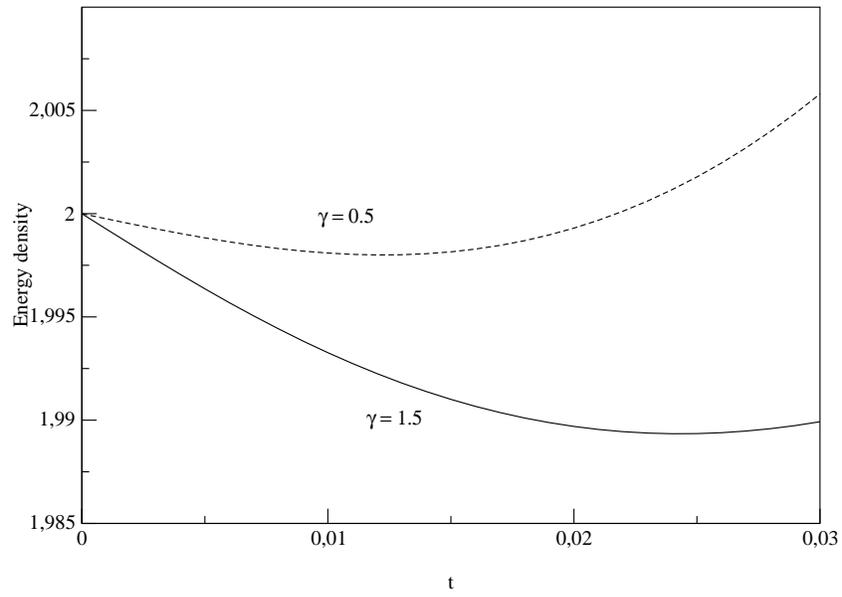}
\caption{Energy density $\epsilon(t)$ vs. time $t$.}
\end{center}
\end{figure}

\begin{figure}
\begin{center}
\includegraphics[width=11cm]{figure6.eps}
\caption{Entropy production rate $\varsigma(t)$ vs. time $t$.}
\end{center}
\end{figure}
\end{document}